\documentclass{aa} % for a referee version
\usepackage{longtable}
\usepackage{lipsum} 
\usepackage{graphicx}
\usepackage{txfonts}
\usepackage{multirow}
\usepackage{hyperref}
\usepackage{microtype}

\usepackage[normalem]{ulem}

\usepackage{natbib}

\makeatletter

\begin{document} 

\title{Expansion and ongoing cosmic ray acceleration in HESS J1731-347}
\titlerunning{Ongoing cosmic ray acceleration in HESS J1731-347}

\author{  V.~Doroshenko\inst{1}\thanks{E-mail: doroshv@astro.uni-tuebingen.de}, 
        G. Pühlhofer\inst{1},
        A. Santangelo\inst{1}
          }
          
\institute{Universit{\"a}t T{\"u}bingen, Institut f{\"u}r Astronomie und Astrophysik T{\"u}bingen, Sand 1, T{\"u}bingen, Germany}

\abstract{
  Diffusive shock acceleration in supernova remnants (SNRs) is considered one of the prime mechanisms for Galactic Cosmic Ray (GCR) acceleration. It is still unclear, however, whether SNRs can contribute to GCR spectrum up to the ``knee'' (1\,PeV) band as acceleration to such energies requires an efficient magnetic field amplification process around the shocks. The presence of such a process is challenging to test observationally. Here we report on the detection of fast variability of the X-ray synchrotron emission from the forward shock in the supernova remnant HESS\,J1731-347, which implies the presence of a strong ($\sim$0.2\,mG) field exceeding background values and thus of effective field amplification. We also report a direct measurement of the high forward shock expansion velocity of 4000-5500\,km/s confirming that the SNR is expanding in a tenuous wind bubble blown by the SNR progenitor, is significantly younger (2.4-9\,kyr) than previously assumed by some authors, and only recently started interaction with the dense material outside the bubble. We finally conclude that there is strong evidence for ongoing hadronic CR acceleration in this SNR.}

\maketitle

\section{Introduction}The identification of the acceleration sites of Galactic cosmic rays (GCRs) up to $\sim 10^{15}$\,eV (PeV) energies remains one of the key open problems of modern high energy astrophysics. Shocks of supernova remnants have been suggested as  plausible acceleration sites \citep{1978ApJ...221L..29B,1994A&A...287..959D,2005JPhG...31R..95H,2010ApJ...718...31P}, and the
broadband emission from X-rays to TeV $\gamma$-rays observed from several supernova remnants (SNRs) firmly establishes this object class as accelerators of electrons up to $\gtrsim$ TeV energies. 
However, for individual objects, it is difficult to prove the existence of accelerated nucleonic 
particles (see \cite{2023arXiv230110257C} for a recent review).
The reason is that TeV $\gamma$-rays can have either leptonic (through Inverse Compton emission from interaction of accelerated electrons with ambient photons) or hadronic origins (through $\pi^0$-decay from interaction of accelerated protons with ambient gas). 
It remains, therefore, an open question which conditions lead to substantial hadronic acceleration in individual SNRs.
The detection of 100 TeV-PeV photons from SNRs has long been considered as key to identify hadronic acceleration in those objects. Recent results \citep{2021Natur.594...33C,2023arXiv230517030C} 
show, however, that the association of such $\gamma$-ray sources to SNRs is not easily possible, and, even more importantly, that the hadronic/leptonic ambiguity prevails even at these photon energies. This might indicate that proving efficient hadronic acceleration in SNRs up to the PeV range using only the $\gamma$-ray channel might be difficult with currently available instrumentation. 

Evidence that efficient hadronic acceleration is indeed ongoing can be obtained, for instance, through the detection of rapid variability of synchrotron X-ray emission from individual shock regions. In \cite{2007Natur.449..576U} the observed rapid variability of extended synchrotron X-ray emission from the SNR RX\,J1713.7-3946 was used to put constraints on the magnetic field.
Indeed, the expected variability timescale of a population of accelerated electrons which are cooled by synchrotron emission is defined by the longer of the two following characteristic timescales \citep{2007Natur.449..576U}: $t_{\rm synch}\sim1.5(B/\rm{mG})^{-1.5}(\varepsilon/\rm{keV})^{-0.5}$ years, the radiative synchrotron cooling timescale of photons with energy $\varepsilon$, and the acceleration timescale $t_{\rm acc}\sim\eta(\varepsilon/\rm{keV})0.5(B/\rm{mG})^{-1.5}(\upsilon_{\rm shock}/3,000 {\rm km/s})^{-2}$ years \citep{2007Natur.449..576U}. The observed X-ray variability in the SNR~RX\,J1713.7-3946 on a year to year time scale \citep{2007Natur.449..576U} implies, therefore, magnetic field strengths of the order of mG so acceleration of nucleonic 
particles is unavoidable as well. 
The observed variability thus strongly supports a hadronic interpretation of the TeV $\gamma$-ray spectrum of RX\,J1713.7-3946, even if the coherent interpretation of other observables probably requires a more complex scenario beyond a simple one-zone emission model \citep{2014MNRAS.445L..70G,2018A&A...612A...6H}. 
Similar variability has already been reported for Cassiopeia~A \citep{2008ApJ...677L.105U}, SNR~G330.2+1 \citep{2018ApJ...868L..21B}, and Tycho's SNR \citep{2020ApJ...894...50O}. While also changes in the magnetic field or turbulence properties have been discussed as possible driver of the variability in some instances, ultimately it was always concluded that magnetic field amplification is the best interpretation of the data.

Here, we add another example and report similar X-ray variability observations from the SNR~HESS\,J1731-347. 
We investigate the variability of the X-ray flux in HESS\,J1731-347 and report two locations of the shock front of the SNR which exhibit variability on a $\sim10$ year timescale. This provides strong evidence for the presence of strong magnetic fields, and thus field amplification in HESS\,J1731-347.  Together with the fact that electron acceleration is definitively ongoing, our result also implies ongoing hadronic acceleration in HESS\,J1731-347 irrespective of interpretation of the observed properties in $\gamma$-ray band. We note that earlier attempts to classify at least some of the TeV emission from this object as hadronic were based on the association with molecular clouds at a distance of $\sim$5.2 kpc \citep{2014ApJ...788...94F} which in meanwhile has been challenged by the robust distance measurement of the central compact object (CCO) associated with the SNR of 3~kpc or below \citep{2022NatAs.tmp..224D,Landshofer21}, although similar arguments can also be made with a gas association at that distance \citep{2017A&A...608A..23D}.
Finally, we also measure the expansion rate of the shell at {900-4000}\,km/s, which implies an age of the SNR in the range of 2-9\,kyr, i.e. significantly lower than an early estimate by \citet{2008ApJ...679L..85T} (30\,kyr) and consistent with more recent estimates 
\citep{2011A&A...531A..81H, 2016A&A...591A..68C, 2016MNRAS.458.2565D}.

\section{Observations and data analysis}
\label{sec:methods}
\begin{figure*}[ht!]
    \centering
    \includegraphics[width=\textwidth]{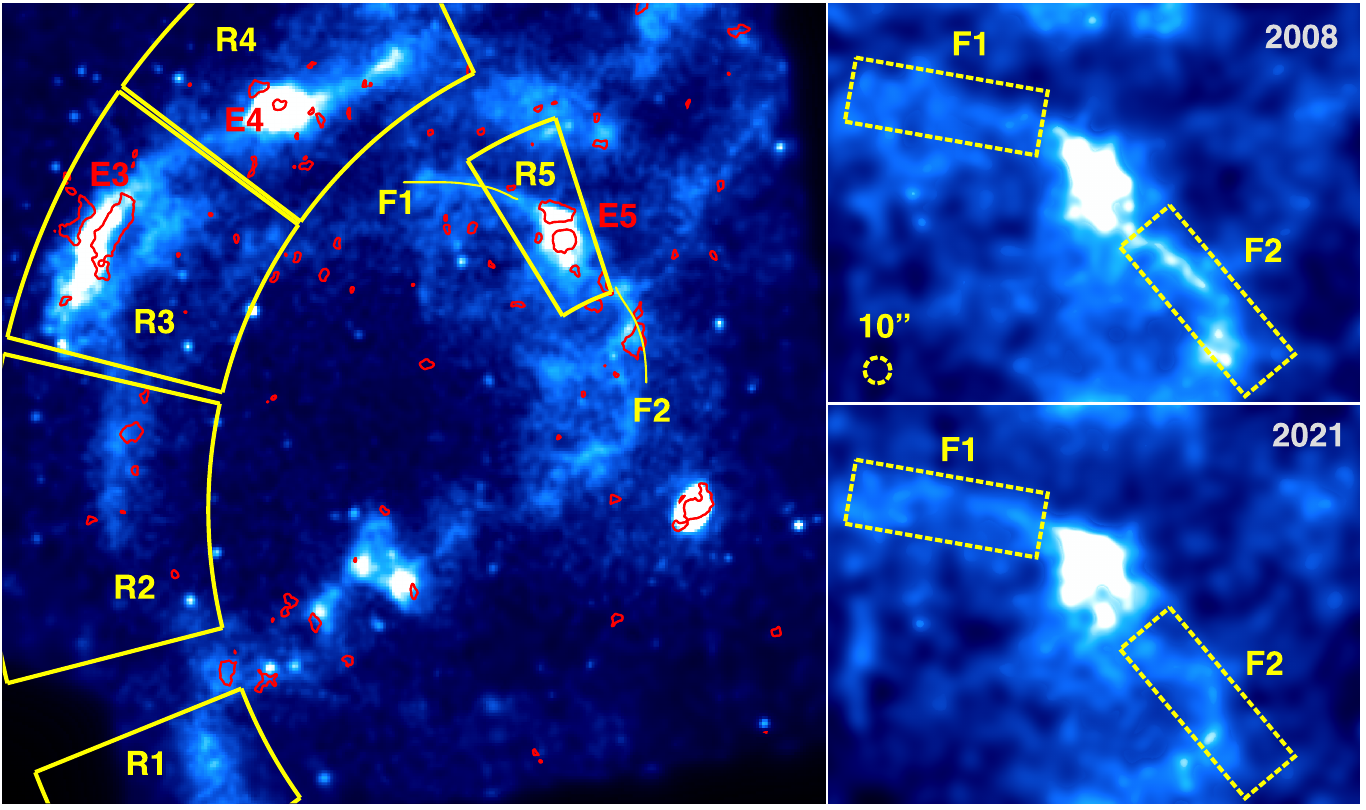}
    \caption{Overview of the \textit{Chandra} data from HESS J1731-347 in the 0.5-7 keV band throughout the mission's lifetime (left). The red contours show regions which were found to exhibit significant variability (3 sigma confidence level) between the 2008 and 2021 observations. 
    The yellow sectors (R1-R5) depict regions that were used to obtain the surface brightness radial profiles used to measure the expansion rate of shell parts. The inset on the right shows a close-up view of a bright knot and two narrow filaments (F1-F2) which exhibit a significant flux variability between the 2008 and 2021 observations, as discussed in the text. All images are adaptively smoothed to contain at least 30 counts per pixel to improve visual appearance and are plotted using a linear scale between 0 and 10$^{-8}$ counts/cm$^2$/s}.
    \label{fig:overview}
\end{figure*}
To assess potential variability and measure the expansion rate of the SNR shell, we rely on high spatial resolution observations of the brightest part of the remnant with \textit{Chandra} \citep{2002PASP..114....1W}. With these goals in mind we proposed an observation of the region with a  configuration that was identical to the earlier observation (obsid 9139),  in order to minimize potential systematic effects associated with the known spatial variations of the point spread function and the effective area over the field of view. The second observation (obsid. 22423) was carried out in 2021, i.e. 13 years after the first one in 2008. The only difference between the two observation set-ups was the increased exposure of the 2nd observation (by 15\%) to compensate for a gradual decrease of the effective area of the telescope due to the accumulation of contaminants on the detector \citep{2020SPIE11444E..97P}. 

Considering the relatively hard power-law spectrum of the extended emission, extending throughout the energy band covered with \textit{Chandra}, a broad energy band (0.5 - 7 keV) was used throughout the analysis to ensure high counting statistics.
Both observations were reduced using the \textit{Chandra} Interactive Analysis of Observations (CIAO)~v.~4.19 following standard procedures described in the instrument's documentation. We also aligned both pointings to a common pixel grid (matching that of the first observation) using the \textit{wcs\_match/wcs\_update} tasks and coordinates of the 33 point sources detected in both observations with the \textit{wavdetect} tool. The residual offset in the alignment between the two observations is estimated to be $\le0.8^{\prime\prime}$ based on the residual scatter of the point source positions detected in the aligned images. This number is smaller than the estimated errors of the measured shifts of the expanding parts of the shell, and much smaller than the absolute values of the largest measured shifts. Nevertheless, the additional systematic error associated with the alignment of the two observations was accounted for in the further analysis. We note that we also reduced all other available observations of the source with \textit{Chandra}, i.e. obs. 21340, 22968, 23146 and 23276, to obtain a higher-quality overview image covering a larger fraction of the remnant, as shown in Fig.~\ref{fig:overview}. The additional observations were not used, however, for the quantitative assessment of the variability due to the significantly lower counting statistics and potential systematic effects related to variation of the point spread function over the field of view, as discussed above.

\subsection{Variability of extended X-ray emission from the shell}
To detect variability from a given sky region, we need to test the hypothesis that the number of counts detected in the two images is inconsistent with being drawn from the same Poisson distribution. Considering that our initial aim is a blind search for variable features in the image (i.e. filaments in the SNR), such a comparison needs to be done for multiple regions probing the relevant spatial scales (i.e. comparable with the width of the filaments), each containing a sufficient number of counts to make the comparison meaningful. 
The latter statement is to some extent ambiguous, and is defined by properties (i.e. algorithms used, power and adopted confidence level) of the statistical test adopted to test the equivalence of the rates from the two regions. 
To our knowledge the most powerful test designed specifically for this purpose is the estimated p-values test (E-test) proposed by \cite{KRISHNAMOORTHY200423} and implemented in the \textit{statsmodels} package \citep{seabold2010statsmodels}. This test requires of the order of a few tens to a few hundreds counts \citep{KRISHNAMOORTHY200423} in each of the samples for a robust comparison of the rates, depending on count numbers and adopted power and confidence level for the test. 
In practice, we selected the angular scale for the variability analysis based on the several considerations. First, the characteristic width of the filaments, i.e. the scale relevant for the variability search, is of the order of 15$^{\prime\prime}$. This coincidentally corresponds to a region size that contains a few tens to a few hundred counts for various parts of the image, i.e. it roughly corresponds to the range quoted in Table~1 of \cite{KRISHNAMOORTHY200423}. Finally, this choice was confirmed empirically, by constructing variability maps as described below for several spatial scales and identifying the smallest scale where random fluctuations still allowed us to visually identify regions of enhanced variability. 

The variability maps were constructed by calculating the number of counts and the total effective exposure, both accumulated within a circle with a fixed radius for each pixel in the aligned images of the two observations. Then, the hypothesis was tested whether the resulting rates match each other, using the E-test as implemented in the \textit{statsmodels.stats.rates.etest\_poisson\_2indep} function.  Here, the known differences in exposure and effective area of the two observations \textbf{is accounted for} by setting the exposure argument of the \textit{etest\_poisson\_2indep} function to the values calculated by the \textit{mkexpmap} task of the \textit{ciao} package. As already mentioned, we repeated the analysis for several spatial scales and found that the initial best estimate of an about 15$^{\prime\prime}$ region size where variability can be expected and can be detected is indeed correct. That is, going below this value results in very noisy variability maps (as expected from the properties of the E-test), whereas increasing the region size starts to smooth out variable features, decreasing their significance. 
Therefore we fixed the probed spatial scale to 15$^{\prime\prime}$ to obtain the final probability map. Here, each pixel encodes the probability that the observed count rates within 15’’ match each other in the two observations. The result is shown with the contours in Fig.~\ref{fig:overview} showing the regions where this hypothesis can be rejected at a $3\sigma$ confidence level and above, i.e. the contours show areas of significant variability. 

In the following, we focus on the relevant variable features around X-ray filaments. 
We note that several small-scale variable regions also visible in the figure (including the region around the central compact object) are mostly associated with point sources and the apparent variability here is associated with the intrinsic variability and residual misalignment between the images of the two observations comparable with the pixel and PSF size.  
That is, masking point sources fully removes this type of variability. On the other hand, several larger regions including a bright filament in the upper left corner of Fig~\ref{fig:overview} labeled R3, a bright knot and two adjacent filaments half way through the shell labeled R5, F1 and F2 stand out and deserve a more detailed discussion.

It is noteworthy that the regions around the outer rim of the shell labeled E3 and E4 as well as at the more inner filament E5, where the extended emission appears to be strongly variable, exhibit a two-peaked structure in the probability maps. This is likely due to the radial expansion of the SNR, as discussed below. Several regions where the emission appears to be intrinsically variable can, however, be identified as well. To confirm that this is indeed the case, we first compare the relative brightness of the two filaments adjacent to the knot R5. Whether variability is detected is not obvious from the presented contours at the chosen confidence level. However it becomes apparent when the images are directly compared, as illustrated in the right panels of Fig~\ref{fig:overview}. That this feature is missed in the blind variability search is largely due to the relatively low brightness of this part of the shell, implying a comparatively low number of counts for the chosen spatial timescale of $\sim15^{\prime\prime}$. Considering the observed morphology of the filaments, the use of larger regions containing more counts is, therefore, justified in this case. In particular, boxes with a size of 50$^{\prime\prime}$x150$^{\prime\prime}$ as indicated in Fig.~\ref{fig:overview} appear to contain most of the flux from both filaments. In total, 966 and 1301 counts in the first, and 1052 and 1037 in the second observations are detected from these regions. The direct application of the E-test to estimate the probability of brightness variability for each of the two regions allows us to reject the hypothesis of equal Poisson rates at $\sim2\sigma$ and $\sim5.5\sigma$ confidence level for the two filaments, i.e. the F2 region dims significantly, while F1 only exhibits statistical fluctuations which is also suggested by the visual inspection of Fig.~\ref{fig:overview}. 

We note that differences in effective exposure, the calibration of the contaminant thickness, and the degradation of the detector could in principle affect our comparison above. However, those effects should affect adjacent extraction regions of the image such as F1 and F2 in a similar way. A flux or count ratio is therefore largely insensitive to instrumental changes and should be the same in both observations if there is no intrinsic variability. This hypothesis, i.e. that the observed count ratio of the first filament is equal to that in the second, can  be statistically tested. To do that we used the same function as for the construction of the variability maps, namely \textit{statsmodels.stats.rates.etest\_poisson\_2indep} to test the hypothesis that the observed number of counts detected in region F1 in two observations is consistent with the hypothesis that their ratio is given by their ratio in region F2 (or vice versa). Considering that the observed count numbers are actually random realizations of a Poisson process, we used a bootstrap method to get a more reliable estimate of the significance. In particular, we repeated the test $10^4$ times, simulating count numbers in each region and each observation from a Poisson distribution with the rate defined by the observed numbers, and tested whether the ratio is the same for the two regions. As a result we found on average a null hypothesis probability of $\sim5\times10^{-5}$ ($\sim4\sigma$ confidence level), i.e. the count ratio between observations of the two filaments is indeed statistically different. 
We conclude, therefore, that the observed relative brightness of filaments F1 and F2 
is significantly variable in the two observations at more than 3$\sigma$ confidence level, even if the conservative procedure comparing rate ratios as outlined above is adopted. The change of relative brightness is likely driven by a change of brightness of filament F2, as follows from the comparison of the count rates from this region in the two observations.
We conclude, therefore, that the observed variability is statistically significant, which is further corroborated by the fact that all identified variable regions appear to be well localized and associated with some of the filaments.

\subsection{Expansion rate of the SNR shell}
\begin{figure}
    \centering
    \includegraphics{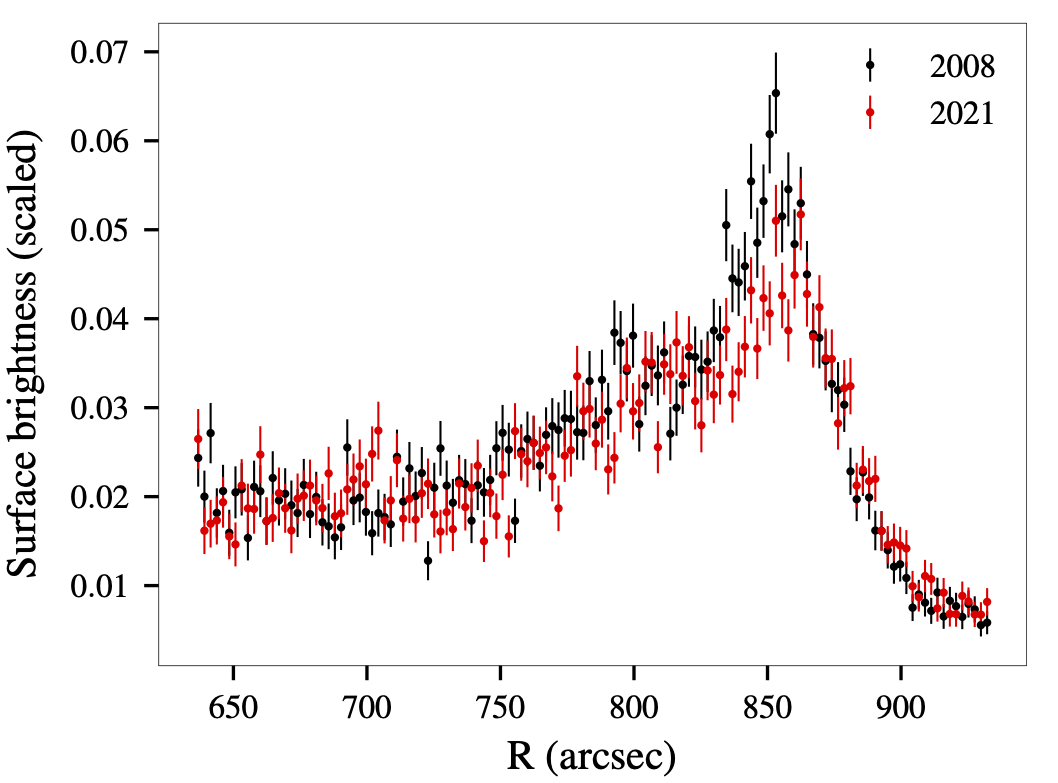}
    \caption{Radial profiles of the X-ray surface brightness at region R3 (defined in Fig. 1) measured in 2008 and 2021 \textit{Chandra} observations, in the 0.5-7 keV energy band.}
    \label{fig:methods_r3}
\end{figure}

\begin{figure}
    \centering
    \includegraphics{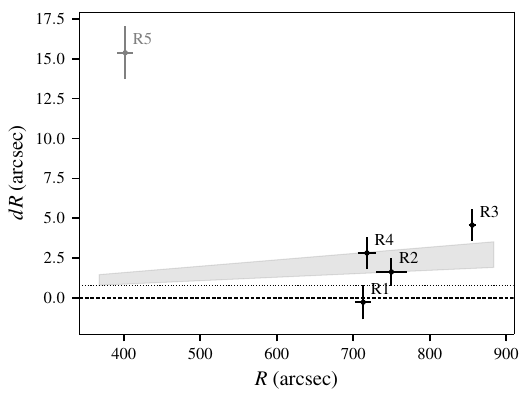}
    \caption{Offsets measured for various parts of the shell (R1-R5) defined in Fig.~\ref{fig:overview}, as described in the text. The grey band shows the 1$\sigma$ expectation for the  offset as a function of distance assuming a uniform expansion as obtained by a joint fitting of the regions R1-R4 (i.e. excluding the outlier region R5). The horizontal lines correspond to zero offset and an 0.8$^{\prime\prime}$ estimated uncertainty, from the alignment of the two observations.}
    \label{fig:m_offsets}
\end{figure}

As already mentioned above, the observed two-peaked patterns in contours reflecting significant variability around the bright  filaments E3, E4 and E5 suggest a positional shift, that is the expansion of the SNR shell rather than intrinsic flux changes. To estimate the expansion rate we constructed partial surface brightness radial profiles around these regions and other parts of the shell containing filament-like structures for both observations. In particular, we integrated the flux within partial annuli with a step size of 2.3$^{\prime\prime}$ in radial direction within regions R1-5 defined in Fig.~\ref{fig:overview}. Representative profiles from region R3 are shown in Fig.~\ref{fig:methods_r3}. A systematic shift between the profiles obtained in 2008 and 2021 is observed, especially at the outer edge of the filament 
Note that besides this shift, the 2008 profile also appears to be more peaked, i.e. the emission morphology appears to have changed, even if it is not trivial in this case to separate the intrinsic variability of the flux from the expansion of the shell. Nevertheless, we attempted to estimate the expansion rate by cross-correlating the radial profiles of the two observations, using a method commonly used to detect phase shifts in radio pulsars as described in \cite{1992RSPTA.341..117T}, which is relatively insensitive to minor morphology changes. The uncertainty of the resulting measurements was estimated using again the bootstrap method. For this, the standard deviation of the shifts was estimated by a cross-correlation of $10^4$ pairs of synthetic radial profiles (for each region) that were simulated based on the observed number of counts in each radial profile distance bin (assuming a Poissonian distribution and separately for each radial bin). We note that the number of counts in individual radial profile bins is relatively high, i.e. $\ge25$ in all cases, so bootstrap biases discussed by \cite{2021A&A...646A..82P} for Poisson sampling are not expected to play a major role. The results are presented in Fig.~\ref{fig:m_offsets} and imply an SNR expansion rate of 0.2-0.36~arcsec/yr (for the shell as a whole, and for the well-defined filament R3 shown in Fig~\ref{fig:overview}).

We emphasize that in the presence of intrinsic variability it is not possible to fully decouple it from shell expansion as a whole, so some caution is needed when interpreting the estimate above. This is illustrated by the large offset measured for region R5 as shown in Fig.~\ref{fig:m_offsets}, which appears to be much larger than expected given its position. This is likely caused by the intrinsic variability visibly affecting the morphology of the extended emission here (see, e.g., the right panels of Fig~\ref{fig:overview}). We interpret the fact that this region appears as an outlier as further indication that not only emission from filaments F1/F2 but also from the knot R5 itself appears to be variable. For this reason we consider any offset value measured for this region unreliable and exclude it from further calculations.

Amongst the examined regions, the largest shift is measured for region R3, where also the outer rim is best defined. Nevertheless, we attempted to estimate the overall expansion rate by fitting all four remaining regions (R1-R4) simultaneously. We assumed a linear expansion, i.e. $dR\propto R$ expected for a spherically expanding shell
and directly fitted the simulated data used to obtain the points in Fig~\ref{fig:m_offsets}. 
In practice, the Bayesian inference package \textit{ultranest} \cite{2021arXiv210109604B} was used to derive parameter values and uncertainties. The likelihood function was defined assuming a normal distribution with $0.8^{\prime\prime}$ width to account for any residual uncertainty from the alignment of the two observations (indicated in Fig.~\ref{fig:m_offsets} with a horizontal line). As a result, we estimate a maximal shift of 2.6(1.1)$^{\prime\prime}$ for the outer radius of 855(5)$^{\prime\prime}$, which translates to an expansion rate of $\sim0.1-0.3^{\prime\prime}$\,yr$^{-1}$ and makes HESS~J1731-347 only the second of the spatially resolved TeV SNR shells (besides RX\,J1713.7-3946, see e.g. \citealt{2016PASJ...68..108T,2017A&A...597A.106A,2020ApJ...900L...5T}) where such a measurement is available.
The kinematic age of the shell can then be estimated at 2.4-9\,kyr, and the shock velocity at 900-4000\,km/s under the assumption of a constant forward shock velocity and a distance of 2.5~kpc \citep{Landshofer21}. These results appear to be consistent with more recent literature estimates \citep{2011A&A...531A..81H, 2016A&A...591A..68C, 2016MNRAS.458.2565D}.

\section{Discussion and conclusions}
Our study is based on the comparison of two X-ray observations of the brightest part of the SNR shell of HESS J1731-347, taken with the \textit{Chandra} observatory in 2008 and 2021, 13 years apart from each other. The second observation was specifically geared towards the search for variability and to measure the expansion rate of the shell, and was therefore set up to fully replicate the former one. This allows a straightforward and robust comparison of both observations to be performed, without introducing systematic effects associated with variations of the point spread function (PSF) and effective area over the field of view of the instrument. In particular, it allows us to conduct a blind search for variability sites by means of a direct comparison of the count-rates between individual observations. We have conducted such a search and constructed a variability map shown with red contours in Fig~\ref{fig:overview}. Several regions with highly significant brightness variations ($\ge3\sigma$ for spatial scales of $\sim15^{\prime\prime}$) between the two observations have been identified. The variability of these regions can either be due to the expansion of the shell or it can be intrinsic.  The variability observed in several regions along the outer rim of the shell appears to be mostly associated with the expansion of the shell, whereas the variability of several filaments within the inner part of the remnant is almost certainly intrinsic. Indeed, the highly significant variability of the relative brightness of the two filaments adjacent to the knot R5/E5 can in no case be explained by expansion of the shell, and thus shows that the intrinsic flux is definitively variable.

We conclude, therefore, that the observed X-ray emission from HESS~J1731-347 is indeed variable on a few years timescale.
Similar variability observed in RX\,J1713.7-3946 has been attributed \citep{2007Natur.449..576U} to ongoing particle acceleration and used to derive a lower limit on the magnetic field strength at the shock. The same arguments can be applied to HESS\,J1731-347, which implies a lower limit of the magnetic field strength of $\sim$0.15-0.19~mG, which strongly supports the hypothesis of efficient ongoing nucleonic acceleration also in this case. We note that variability can in principle be also related changes in magnetic field strength or magnetic turbulence, so the detection of variability itself can not be considered a smoking gun evidence for ongoing nucleonic acceleration. On the other hand, for HESS\,J1731-347 the highly significant difference of the azimuthal dependence of the X-ray/radio and TeV surface brightness \citep{2017A&A...608A..23D} has already been used to argue for at least a partly hadronic origin of the TeV emission in the X-ray dim Western part of the shell.
More importantly, the observed variability is confined to the opposite part of the shell, which indicates that the acceleration is currently ongoing throughout large parts of the shell.

We observe that HESS\,J1731-347 appears to have similar characteristics as RX\,J1713.7-3946, besides the X-ray variability detection.
Both are remnants of core-collapse supernovae which exploded in a wind bubble blown by the progenitor, and only recently did their ejecta begin to interact with a dense ISM at the edge of the bubble \citep{2019ApJ...887...47C,2017A&A...608A..23D,2021ApJ...915...84F}. 
For both sources, expansion in a rarefied interstellar medium can explain the still high shock velocities, despite their large physical sizes of several tens of parsecs and ages of several thousand years. High expansion speed and high ISM densities are, in fact, essential ingredients for acceleration up to PeV energies (e.g. Eq 7 in \citealt{2021Univ....7..324C}) and are not necessarily easily combined. 
Expansion in a wind-blown bubble is a natural scenario to produce such a combination \citep{2016A&A...595A..33T}, and the fact that both of these SNRs appear to follow it suggests that it might not be uncommon for efficient hadronic acceleration. 
In this scenario, one can expect the duration of the acceleration period to be relatively short, as the shock is expected to quickly lose most of its energy upon reaching the dense bubble walls.  Indeed, we note that the observed circular shape of HESS\,J1731-347 in coincidence with dense material around the western rim \citep{2017A&A...608A..23D} implies that the interaction with material outside the bubble has only started recently. 
This, and the need to have a bubble in the first place, might explain the low observed number of SNRs where hadronic acceleration to $\gg$TeV energies is plausible. Finally, we note that our findings also agree with the conclusion that CR acceleration occurs in environments enriched by stellar outflows to explain the observed overabundance of 22Ne/20Ne \citep{2020MNRAS.493.3159G}.

Finally, as discussed above, the measurement of the expansion rate allows us to robustly constrain the age of the remnant, resolving the tension between earlier estimates of the age of around 30~kyr \citep{2008ApJ...679L..85T}
and constraints on the cooling of the neutron star inside the SNR  \citep{2015A&A...573A..53K,2015MNRAS.454.2668O}.
Our results thus strongly support the suggestion by \cite{2020MNRAS.496.5052P} that a much lower age of 2-6\,kyr is more plausible in context of the observed properties of the central neutron star (and, in fact, consistent with several revised estimates of the SNR age \citealt{2011A&A...531A..81H, 2016A&A...591A..68C, 2016MNRAS.458.2565D}). We note in passing that the revised temperature and luminosity of the CCO \citep{2022NatAs.tmp..224D} had already further reduced this tension.

\begin{acknowledgements}
The scientific results reported in this article are based to a significant degree on observations made by the \textit{Chandra} X-ray Observatory and made use of software provided by the \textit{Chandra} X-ray Center (CXC).
\end{acknowledgements}

\bibliographystyle{aa}
\bibliography{bibtex.bib}

\end{document}